\documentclass[11pt]{article}
\usepackage{graphicx}
\usepackage{amssymb}

\title{ Band-gap structure and  singular character of  bounded
one-dimensional  
multibarrier potential }
\author{D. Bar}

\begin{document}
\maketitle

\bigskip

\begin{abstract}   \noindent 
{\it  
  The bounded 
one dimensional multibarrier potential shows signs of chaos,  phase 
transition and a transmission
probability of unity for 
certain values of its total length $L$ and the ratio $c$  of   total interval to 
total width.      Like the infinite Kronig-Penney system, 
which is arranged along the whole spatial region,  the bounded multibarrier
potential  has   a band-gap
structure in  its energy spectrum. But unlike the Kronig-Penney 
system,  in which   the gaps disappear for large  energies,  
  these gaps
  do not disappear  for certain values of $L$ and $c$.  
   The energy  is discontinuous 
  even in  parts of the spectrum with  no gaps at all.  These results 
    imply 
that  the  
energy spectrum of the bounded  multibarrier system  
is singular.  }
\end{abstract}
\noindent
{{\bf keywords}: Band-gap structure, Multibarrier potential, Singular spectrum}

\bigskip

\bigskip \noindent 

\protect \section{Introduction \label{sec1}}

It is known that the Kronig-Penney multibarrier system \cite{Merzbacher,Tannoudji} 
   which is arranged along the whole spatial region is characterized by an energy
   spectrum which  is composed of continuous bands separated by forbidden gaps
   \cite{Merzbacher,Tannoudji,Kittel}. The larger are the values of the energies 
    the shorter these gaps become until they entirely disappear for
   large enough energies \cite{Merzbacher,Tannoudji,Kittel}. 
It is also known \cite{Merzbacher} that the gaps in the energy spectrum of the
infinite Kronig-Penney system are related 
 \cite{Merzbacher}  to the eigenvalues $\lambda_{\pm}$ of the characteristic
 equation which  is
 associated with the two dimensional transfer matrix that relates the
 transmission and reflection coefficients of any two neighbouring barriers.  
 The limits ($\lim_{n \to \pm \infty}\lambda^n_{\pm}$) 
 are 
 taken   when the related potential barriers are separated from each other
 by a very large number $n$ of other similar barriers.  Thus, if either of these 
 limits  becomes very  large, in which case the resulting wave function
 does not remain finite,   the corresponding energies are disallowed
 \cite{Merzbacher} and these constitute the gaps of the energy spectrum. \par 
 The former  problems which  are associated with the infinite Kronig-Penney
 system do not arise in the bounded  multibarrier system
 discussed here. This is because although the number of barriers is very large, 
 as for the Kronig-Penney system,  nevertheless the finite extent of the system
 enables one to analytically express in explicit  form \cite{Bar1,Bar2}   
 the total transfer matrix that relates the two barriers at the two 
 extreme sides of the  system. Thus, the characteristic equation and the
 corresponding eigenvalues of this overall transfer matrix 
  have also been found \cite{Bar3} in exact closed forms. 
    From the latter  one may
  construct the appropriate energy spectrum for both cases of 
  $E\!>\!V$ and $E\!<\!V$  
  where $E$ is the energy and $V$ the constant height of all the barriers. 
  \par 
  An important parameter which will be constantly referred  to in this work is
  related to the mentioned eigenvalues $\lambda_{\pm}$ and is the variable over
  which the energy $E$ is plotted (see, for example, Figure 6.6 in
  \cite{Merzbacher} and the seven figures here). This parameter is denoted
  $\kappa$ here and is defined by the first of the following Eqs (\ref{e8}) (see
  also Eqs (\ref{e9})-(\ref{e10})). 
  We show in the following  that  there are no  allowed energies 
  that correspond to    certain   values of the
  total length $L$ of the system and the ratio $c$ of its total interval to total
  width.     
  Moreover, in contrast to the Kronig-Penney system in which the gaps in the
  energy spectrum disappear \cite{Merzbacher} for either large values of the 
  energy $E$ or (and) large   values of the  parameter  $\kappa$ 
  \cite{Merzbacher}  
         the case here is different. 
  This is because, as will be shown in Sections 3-4,  the gaps in the 
  energy spectrum of the
  bounded one-dimensional multibarrier system do not disappear for certain
  values of $L$ and $c$  even for  large $E$ or (and) large $\kappa$. 
   We also show that the energy $E$ depends   upon the total length
  $L$ and 
  the ratio $c$ in such a manner  that its form, as function of $\kappa$, 
   is quasi-periodic for small $c$ and large (or  intermediate) values of  
    $L$  
  and  is proportional to $\kappa$ for all  $c>3$. It will   also
  be shown  that the energy spectrum is discontinuous at many of its points 
   even at the parts at which it has no gaps at all. \par
  In  Section 2 we introduce the terminology and terms \cite{Bar1} that
   describe the bounded multibarrier potential. In
  Section 3 we discuss the energy spectrum for the $E\!>\!V$ case and find the
  values of the parameter $\kappa$ for which no corresponding energies are
  found. That is, we find the band-gap structure of it. We also show the
  remarked discontinuity in $E$  as function of $\kappa$. In Section 4 we repeat
  the whole  process for the $E\!<\!V$ case and summarize in Table 1 the allowed
  energies for both cases of $E\!>\!V$ and $E\!<\!V$. In Section 5 we 
  show that  the occurence of  stable   gaps in its energy spectrum  
   together with other  previously shown  \cite{Bar1,Bar2,Bar3}  
  properties of  it, imply 
  that   the bounded multibarrier 
  potential  is a singular system 
  \cite{Reed,Grossmann,Pearson,Avron,Demkov,Jitomirskaya,Carmona}.  
   We conclude with a brief summary in Section
   6.     
\bigskip \noindent 

\protect \section{The bounded one-dimensional multibarrier  potential   \label{sec2}}
We consider a  bounded one-dimensional system of  $N$ barriers  
 assumed to have the same height $V$ for all of them  and to be uniformly
 arrayed  between  the point $x=-\frac{a+b}{2}$ 
and $x=\frac{a+b}{2}$. Thus,   the total length of the system is $L=a+b$  
where  $a$ denotes  the total width of all the $N$ barriers (where the potential 
$V \ne 0$), and $b$
is the total sum of all the intervals between neighbouring barriers 
(where $V = 0$). That is,  in a system of $N$ barriers 
the width of each one  
 is  $\frac{a}{N}$, and the interval between each two neighbours   is 
$\frac{b}{N-1}$. Writing  $b$ as $b=ac$,  where $c$ is a real number, 
 we can express $a$
and $b$ in terms of $L$ and $c$ as  \cite{Bar1} 
\begin{equation} \label{e1} a=\frac{L}{1+c}, \; \; \; \; \; 
 b=\frac{Lc}{1+c} \end{equation}  
We  consider the passage of a plane wave through this system, which has
the form $\phi=A_0e^{ikx}+B_0e^{-ikx}$ where  $ x \le  -\frac{a+b}{2}$. 
Matching boundary conditions at the beginning and end of each barrier,  we may 
construct a solution in terms of the transfer matrix 
\cite{Merzbacher,Tannoudji,Yu}  $P^{(j)}$ on the $j$-th
barrier. Thus,  using the terminology in
\cite{Merzbacher},  we obtain after the $n$-th barrier 
the following transfer matrices equation  \cite{Bar1}
 \begin{equation} \label{e2} \left[ \begin{array}{c} A_{2n+1} \\ B_{2n+1} 
\end{array} \right] = P^{(n)}P^{(n-1)}\ldots P^{(2)}P^{(1)}
\left[ \begin{array}{c} A_0\\ B_0 \end{array} \right]=\cal P 
\left[ \begin{array}{c} A_0\\ B_0 \end{array} \right],  
\end{equation}
where $A_{2n+1}$ and $B_{2n+1}$  are the amplitudes of the 
transmitted and reflected 
parts \cite{Bar1} respectively of 
the wave function at the $n$-th  barrier. $A_{0}$  is the coefficient 
of the initial wave that 
approaches the potential barrier system, and $B_{0}$  is the coefficient 
of the reflected wave from 
the first barrier. $\cal P$  is the total transfer matrix over the  
system of $n$ barriers and is given  for the $E\!>\!V$  case in the limit of 
 very large $n$  
  by  
  \cite{Bar1}  
\begin{equation} \label{e3} {\cal P}_{E\!>\!V}=\left[ \begin{array}{c c}
e^{-iz}(\cos{\phi}+if\frac{\sin(\phi)}{\phi}) &ie^{-iz}d\frac{\sin(\phi)}{\phi} 
\\ -e^{iz}d\frac{\sin(\phi)}{\phi}&e^{iz}(\cos{\phi}-if\frac{\sin(\phi)}{\phi})
\end{array} \right] \end{equation} 
The parameters  
$f$, $d$, $z$ and $\phi$ are expressed as  
\cite{Bar1}  \begin{equation} \label{e4} 
 f=kb+aq\frac{\xi}{2},  \ \ \ d=aq\frac{\eta}{2}, \ \ \ z=k(a+b), \ \ \ 
 \phi=\sqrt{f^2-d^2},  \end{equation} where $k$,  $q$, $\xi$ and $\eta$  are  
 \cite{Bar1} 
\begin{equation} \label{e5} k=\sqrt{\frac{2m(E)}{\hbar^2}}, \ \ \  
q=\sqrt{\frac{2m(E-V)}{\hbar^2}}, 
\ \ \ \xi=\frac{q}{k}+\frac{k}{q}, \ \ \  
  \eta=\frac{q}{k}-\frac{k}{q}  \end{equation}  
  It is shown \cite{Bar1} that $\cal P$ 
  is given for the  $E\!<\!V$ case and in the same limit of very large $n$  by 
   \begin{equation} \label{e6} 
  {\cal P}_{E\!<\!V}=\left[ \begin{array}{c c}
e^{-iz}(\cos{\grave \phi}+\frac{i\grave f\sin(\grave \phi)}{\grave \phi}) 
&-ie^{-iz}\frac{\grave d\sin(\grave \phi)}{\grave \phi} 
\\ ie^{iz}\frac{\grave d\sin(\grave \phi)}{\grave \phi}&
e^{iz}(\cos{\grave \phi}-
i\frac{\grave f\sin(\grave \phi)}{\grave \phi})
\end{array} \right],  \end{equation} 
 where the parameters  $\grave f$,  $\grave d$ and $\grave \phi$ are now    
   \begin{equation} \label{e7} \grave f=kb-\frac{aq\eta}{2}, \ \ \   
\grave d=\frac{aq\xi}{2}, \ \ \  \grave \phi=
\sqrt{\grave f^2-\grave d^2},  \end{equation}    $k$ is the same as for the
$E\!>\!V$ case and 
 $ q= \sqrt{\frac{2m(V-E)}{\hbar^2}}$.  
 In the numerical part of this work we follow the convention in the literature 
 that 
 assigns to   $\hbar$ and $m$ the values of 1 and $\frac{1}{2}$ respectively.  
Defining, as in \cite{Bar3},  the parameters $\kappa$ and $\tau$  
\begin{equation} \label{e8} e^{i\kappa}=\frac{\cos(\phi)+
i\frac{f\sin(\phi)}{\phi}}{\sqrt{\cos^2(\phi)+\frac{f^2\sin^2(\phi)}{\phi^2}}}, 
\ \ \ \  \tau=1+\frac{d^2\sin^2(\phi)}{\phi^2}, \end{equation} 
 one may obtain the 
eigenvalues of
either the matrix of Eq (\ref{e3}) for the $E\!>\!V$ case or of the matrix 
(\ref{e6}) 
for  $E\!<\!V$ in the form 
\begin{equation} \label{e9}  \lambda_{1,2}=\tau \cos(\phi-\kappa) 
\pm \sqrt{\tau^2 \cos^2(\phi-\kappa)-1} \end{equation}  The eigenvalues of the
$E\!>\!V$ case are obtained by substituting the  $\phi$ and $f$   
 from  Eq (\ref{e4})  and those 
of the $E\!<\!V$ case by substituting the corresponding quantities  from  
 Eq (\ref{e7}). \par 
  We, now, show    that,
   unlike the infinite  Kronig-Penney system 
   \cite{Merzbacher,Tannoudji,Kittel} that have
   no finite  total length $L$ and no finite ratio $c$, 
   the energy spectrum of  the bounded multibarrier potential 
   depends critically upon  $L$  and  $c$.
     We first  take  the real components of both sides of 
    the first of Eqs (\ref{e8})  and divide  the numerator and denominator of 
    its right hand side by $\cos(\phi)$
    \begin{equation} \label{e10} \cos(\kappa)= 
    \frac{1}{\sqrt{1+\frac{f^2\tan^2(\phi)}{\phi^2}}} \end{equation}
    We note that the last equation is valid for both cases of $E\!>\!V$ 
    and $E\!<\!V$
    except that one should substitute the appropriate $f$ and $\phi$ from Eq
    (\ref{e4}) for the $E\!>\!V$ case (as we do in next section)  
    or $\grave f$ and $\grave \phi$ from Eq (\ref{e7}) for $E\!<\!V$ as 
    done in Section 4.
  
\bigskip \noindent 
\protect \section{The band-gap structure of the finite multibarrier system 
   for the $E\!>\!V$ case\label{sec3}} 
     We, now,  consider   the $E\!>\!V$ case and
     substitute  in Eq (\ref{e10}) the appropriate   $f$, $d$, $\phi$, $k$  
     and $q$  
     from  Eqs (\ref{e4}) and (\ref{e5})   to  obtain 
     \begin{eqnarray} && \cos(\kappa)=
     \frac{1}{\sqrt{1+(1+\frac{a^2V^2}{4E(E(a+b)^2-V(a^2+ab))})
     \tan^2(E(a+b)^2-V(a^2+ab))}}=  \nonumber \\ && = \frac{1}
     {\sqrt{1+(1+\frac{V^2}{4E(1+c)(E(1+c)-V)})
     \tan^2(L^2(E-\frac{V}{1+c}))}} \label{e11} \end{eqnarray}
     The last result was obtained by expressing $a$ and $b$ in terms of $L$ and
     $c$   according to Eq (\ref{e1}). Squaring both sides of the last equation
     one obtains \begin{equation} \label{e12} \cos^2(\kappa)=\frac{1}
     {1+(1+\frac{V^2}{4E(1+c)(E(1+c)-V)})
     \tan^2(L^2(E-\frac{V}{1+c}))}  \end{equation}
     From Eq (\ref{e12})  we see  that for all values of $\kappa$ that
     cause $\cos^2(\kappa)$ on its  left hand side  to vanish
     one must have corresponding values of $L^2(E-\frac{V}{1+c})$ that cause 
     $\tan^2(L^2(E-\frac{V}{1+c}))$ on its right hand side to become 
     very large. That is, for 
     $\kappa=\pm \frac{(2N+1)\pi}{2}$, $(N=0, 1, 2, 3, \ldots)$ one must  have 
     $L^2(E-\frac{V}{1+c})= \frac{(2N+1)\pi}{2}$, 
     $(N=0, 1, 2, 3, \ldots)$.  Note that since $E$, $V$ and $c$ are positive 
       the expression 
     $L^2(E-\frac{V}{1+c})$ for the $E\!>\!V$ case is positive.  Thus, 
      the appropriate values of the energies $E$ that correspond to 
      the remarked
        values of $\kappa$ are $E= \frac{(2N+1)\pi}{2L^2}+\frac{V}{1+c}$.  
	That is,  from the last
     expression for $E$ and from the condition $E\!>\!V$  we find that 
     there are allowed
     energies only for those values of the total length $L$ and the ratio $c$
     that satisfy the inequality \begin{equation} \label{e13} 
     \frac{(2N+1)(1+c)\pi}{2cL^2} > V, \ \ \ N=0, 1, 2, \ldots \end{equation} 
     That is, the energies that correspond to 
     $\kappa=\pm \frac{(2N+1)\pi}{2}$, $(N=0, 1, 2, 3, \ldots)$
   depend upon the total length $L$ of the system and the ratio $c$ of its 
  total interval to total width.
     For all
     other values of $\kappa$ we may take the reciprocals of both sides of Eq 
     (\ref{e12}) and subtract 1 from the resulting expressions   to obtain 
     \begin{equation} \label{e14} \tan^2(\kappa)= (1+\frac{V^2}{4E(1+c)
     (E(1+c)-V)})
     \tan^2(L^2(E-\frac{V}{1+c}))  \end{equation}
     We note that exactly the same equations  as those of 
      (\ref{e12}) and 
     (\ref{e14}) 
     are obtained also for the $E\!<\!V$ case except that  we use, as remarked, 
     the $\grave f$ and  $\grave \phi$ from Eqs (\ref{e7}) 
     with $ q= \sqrt{\frac{2m(V-E)}{\hbar^2}}$ (where, for numerical purposes,    
  we assign, as noted,  $\hbar=1$, and $m=\frac{1}{2}$). From Eq (\ref{e14}) we
  may find the energies which correspond to the values of $\kappa$ that cause
  $\tan^2(\kappa)$ on its left hand side to vanish. These  $\kappa$'s 
  are $\kappa=\pm N\pi$, ($N=0, 1, 2, 3 \ldots$) so    using  Eq
  (\ref{e14}) and the positiveness of $L^2(E-\frac{V}{1+c})$  we  have 
  $L^2(E-\frac{V}{1+c})= N\pi$  from which one obtains   
  $E= \frac{N\pi}{L^2}+\frac{V}{1+c}$.    From the last
     expression for $E$ and from  $E\!>\!V$  we find that there are allowed
     energies that correspond to $\kappa=\pm N\pi$, ($N=0, 1, 2, 3 \ldots$) 
     only for those values of  $L$ and  $c$
     that satisfy   
  \begin{equation} \label{e15}
   \frac{N(1+c)\pi }{cL^2} > V, \ \ \ N=0, 1, 2, \ldots \end{equation} That 
   is, as for the case of 
   the inequality   (\ref{e13}),  the 
  energies that correspond to $\kappa=\pm N\pi$, ($N=0, 1, 2, 3 \ldots$) 
  depend upon the total length $L$ of the system and  the ratio $c$ of its 
  total interval to total width. 
    Note that
  although $L^2(E-\frac{V}{1+c})=0$ corresponds  also to  the former 
  values of $\kappa$
  that result in  
  $\tan^2(\kappa)=0$ there is no value of the energy $E$, for the $E\!>\!V$ case, 
  that corresponds to  $L^2(E-\frac{V}{1+c})=0$.  This is because from 
   $L^2(E-\frac{V}{1+c})=0$ one obtains $E=\frac{V}{1+c}$ which, for positive
   $c$, does not conform to $E\!>\!V$. 
   \par
  Now,  as may be 
  seen from Eq (\ref{e14}), the dependence of the allowed energies   upon 
  $L$, $c$ and $N$ is not restricted  only to these values of $E$  
  that correspond to 
  $\kappa=\pm \frac{(2N+1)\pi}{2}$ or to 
   $\kappa=\pm N\pi$, ($N=0, 1, 2, 3 \ldots$). Moreover, as will be shown 
   in the following,  the form   of the allowed 
      energy, as function of   $\kappa$,  
      depends in a peculiar manner  upon  the ratio $c$. 
       That is, if  $c$ is
      small, say $0<c<3$,   then the expression that multiplies
      $\tan^2(L^2(E-\frac{V}{1+c}))$ on the right hand side of Eq (\ref{e14}) 
      is  larger than unity and  the solution of
        it  for the allowed energy $E$  
   as function  of 
   $\kappa$ can  be obtained  only  numerically. In such case   we find, as shown in the
   following, that these $E$'s change periodically   
      with $\kappa$. If, on the other hand, $c \ge 3$  then
      the expression that multiply $\tan^2(L^2(E-\frac{V}{1+c}))$ on the right
      hand side of Eq (\ref{e14}) is very close to unity. 
      For example, for $c=3$  and $E \approx V$
      this expression equals $1.020$   and it tends fastly to
      unity for  values of $c>3$  or$/$and  $E\!>\!V$. Thus, for $c>3$ we may 
      safely approximate  Eq
      (\ref{e14}) as  \begin{equation} \label{e16} \tan^2(\kappa) \approx  
     \tan^2(L^2(E-\frac{V}{1+c})),   \end{equation}
    from which we obtain $\tan(L^2(E-\frac{V}{1+c}))= \pm \tan(\kappa)$. 
     The  last equation results in   
    $L^2(E-\frac{V}{1+c})= N\pi \pm \kappa$,  \ $(N=0, 1, 2, 3, \ldots)$  
    from which we have    
    \begin{equation} \label{e17} 
    E= \frac{N\pi \pm \kappa}{L^2}+\frac{V}{1+c}, 
    \ \ \ \ \ (N=0, 1, 2, 3, \ldots)
    \end{equation} 
    As seen from Eq (\ref{e17})  not all 
     values of $N\pi \pm \kappa$ are suitable for the $E\!>\!V$ case but only those
    that satisfy $(N\pi \pm \kappa)>\frac{VL^2c}{1+c}$.  For example, for $V=L=15$ 
and $c=3$ the permissible
    values of $N\pi \pm \kappa$ that have 
    corresponding energies are only those that
    satisfy $(N\pi \pm \kappa) > 2531.25$. \par
    For other values of $c$ from the range $0 <c < 3$ the expression that
    multiplies $\tan^2(L^2(E-\frac{V}{1+c}))$ on the right hand side of 
     Eq (\ref{e14}) is, as remarked,
    larger than unity for the   $E\!>\!V$ case    
    and one must resort to numerical methods
    for solving Eq (\ref{e14}) for $E$ as function of $\kappa$. 
    In this case  the
    obtained energies are  {\it quasi-periodic}  where the character and form 
     of the quasi-period
      depend upon the total length $L$ of the
    system  and the ratio $c$ of its total interval to total width.  We have
    emphasized the word quasiperiod since as one may realize from a close
    scrutiny of the appended figures the forms of the energy as functions of
    $\kappa$ are not exactly repeated over the $\kappa$ axis but differ in the
    small details. 
    It is also shown for both cases of  $E\!>\!V$ and $E\!<\!V$  that the
     quasi-periodic character  of  the energy as function of $\kappa$ emerges  
    only for values of $E$ that are close to the potential $V$ (close from below for
    the  $E\!<\!V$ case and from above for $E\!>\!V$). 
      For example, for  $E\!>\!V$
and  $V=15$ the corresponding  energy   is  found to vary in almost 
periodic manner   only 
 between $E=15.5$ and $E=19.5$. This is because 
    for higher values of $E$ the expression that multiplies  the {\it tangent} 
     function  
    on the right hand side of Eq (\ref{e14}) tends to unity even for very small
    values of $c$. In such case the situation is the same as 
    that formerly found
    for  large  $c$  ($c>3$) where  the appropriate
    equation to use is  (\ref{e16}) which  leads to the  linear
    expression (\ref{e17})  for  the energy  as function of $\kappa$.  
      For  values of $E$ that are close to the potential $V$ one may obtain not
     only the remarked quasi-periodic forms of the energy as function of $\kappa$ but
     for very small  $c$ the allowed energies become constant as 
     demonstrated by  the horizontal lines  in  Figures 1-2.      
    This is because  for $E \approx v$ and $c \approx 0$ the right hand side of Eq
    (\ref{e14}) becomes  the indeterminate form $\frac{0}{0}$, so using L'hospital theorem
    \cite{Pipes} we obtain for it the value of 1. In such case the value 
    of $E \approx V$ does not change with $\kappa$. \par
     The quasi-periodic nature  of the energy as function of $\kappa$  
     is  demonstrated for
    small  $c$ in the form of half squares as seen from Figure 1 which shows nine different curves 
    of $E$ from Eq (\ref{e14}) as function of $\kappa$.  
    All the nine curves are  drawn for the
    same values of $V=15$ and  $L=100$ but for 9 different
    values of the ratio  $c=0.2\cdot n$,  $(n=1, 2, 3, ...9)$.  The different
    graphs  in this figure and in  Figures 2-3  correspond to the
    values of $c$ in a such a  manner  that  the  curves with the most
    pronounced  half squares, which are  shown at 
      the upper part of these figures,   fit the
    lower values of $c$.  The curves which are characterized by a 
     short and flat  half squares, which are generally located at the
     lower parts of these figures,  
     correspond to the higher values of $c$.   The
    horizontal lines at the bottom  of Figures 1 and 2 correspond, as remarked, 
      to the constant value of the energy $E \approx V$.  
      The remarked  correspondence between the graphs of $E$ and the
     appropriate $c$'s,  by which the  graphs at the upper part of the figure 
     fit the lower $c$'s and those at the lower part of it fit the higher $c$'s, 
        applies also   for the $E\!<\!V$ case in Figures 4-6. \par
        As seen from Figure 1  
    any two neighbouring half squares   are connected to each other by short horizontal
    lines where the energy is clearly discontinuous at the connecting points.
    That is, at these points the energy jumps in a discontinuous manner to form
    the squared parts where the smaller is $c$ 
    the larger are the corresponding jumps  as may be seen at the
    higher part of the figure.  For growing  $c$  the jumps become smaller  
     and the
    corresponding  half squares shorter   as  seen at  
    the lower part of  Figure 1.   One may also realize that 
    for the  large
    value of $L$ used in this figure the 9 different curves are shown mixed 
    inside each  other between the  two values of  $E \approx 15.3$ and 
    $E \approx 17.8$. Plotting the energies  $E$ as functions of $\kappa$ for  
     additional larger  values of $c$, while keeping the same  
 $L$ as before,  add more  periodic forms (not shown) 
that are arrayed inside the formers between the same two  limits 
of Figure 1.  \par 
We must note that taking into account  the periodic character  of the 
{\it tangent} function we can rewrite Eq (\ref{e14}) in its most 
general form
as  \begin{equation} \label{e18} \tan^2(\kappa \pm N\pi)= (1+\frac{V^2}{4E(1+c)
     (E(1+c)-V)})
     \tan^2(L^2(E-\frac{V}{1+c}) \pm {\grave N\pi}),  \end{equation}  
     where $N, {\grave N}=0, 1, 2, \ldots$.  Since the 
     parameter $L$ occurs only
     under the {\it tangent} function at the right hand side of Eq (\ref{e18}) 
       we may numerically solve 
     it for $E$  and find the same  
      quasi-periodic dependence upon $\kappa$ for different values of $L$. 
            For example, Figure 4, which was drawn for $L=30$ may be
      obtained in almost the same form also for $L=110$.  Figures 1-6 
      of  this  work 
      are obtained  from the equivalent equation  (\ref{e14}) in 
       which ${\grave N}, N=0$. Note that for very small $L$ one may approximate
       the {\it tangent} function at the right hand side of Eq (\ref{e14}) by 
       $L^2(E-\frac{V}{1+c})$ and numerically solves the resulting equation 
       for $E$ as function of $\kappa$. The  spectrum obtained is shown in
       Figure 3 for the $E\!>\!V$ case and in Figures 6-7 for  $E\!<\!V$.  \par 
As $L$
    decreases  the different curves begin to be separated from each other
    and to occupy different sections of the ordinate axis.   
    This is  shown in Figures 2
    and 3 which are drawn under exactly the same conditions and 
    for the same values of
    $V$ and  $c$ as in Figure 1 except that now $L$ assumes the values of $L=5$ 
  for  the 
    curves of Figure 2 and $L=0.3$ for those of Figure 3. 
    The highest  half squares   of Figure  2 
are drawn around  the value of $E=19.6$ and those of Figure 3  around $E=29$.
Note that the half squares  of Figure 2 are each connected to the horizontal
linear sections at more than 
 two
 points (compare with Figure 1). This is  
especially demonstrated in the curves at the higher part of  Figure 2 which 
correspond to the smaller
values of $c$. Also, as in Figure 1 the discontinuous jumps in the 
energy 
are larger for the curves that correspond to   small $c$'s  
 at the upper part of  
Figure 2 
and   smaller for the larger   $c$'s  at the lower part of it.  
The almost periodic curves of  Figure 3 which are all drawn for 
the  value of
$L=0.3$ are seen to be rather curved than squared  and  the  points at which 
 the energy jumps to form the pronounced  parts of it  are  two for each  
 quasi-period 
 as in Figure 1. 
  Also, the larger is the value of $c$ at the lower part of Figure 3 
  the smaller become the  corresponding jumps of $E$.

     \begin{figure}
     \includegraphics{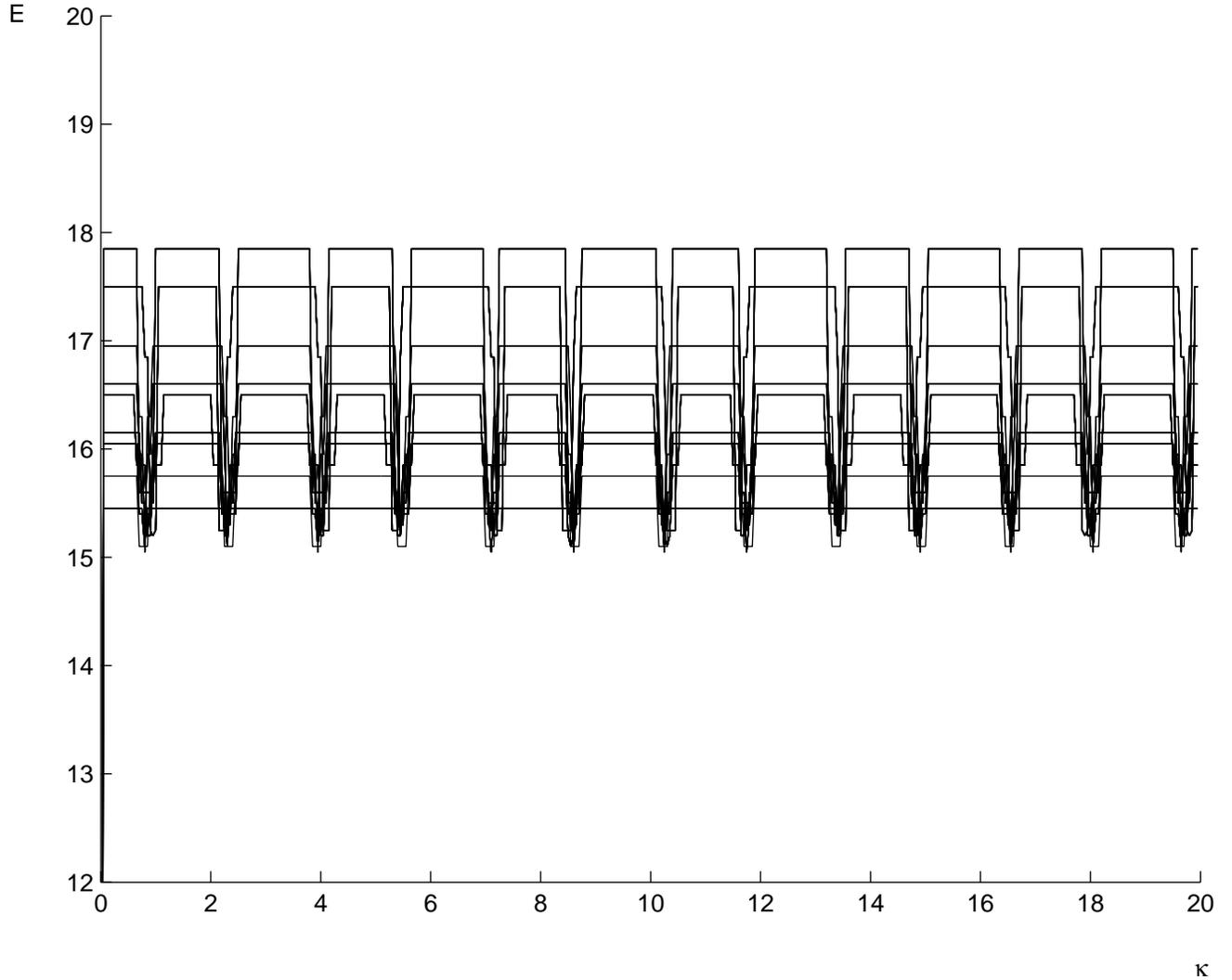}
     \caption{The nine graphs of this figure, which show the energy $E$ from Eq
     (\ref{e14}) as function
     of $\kappa$,   are all drawn for the same values of
$V=15$ and $L=100$ and for nine different values of the ratio  
$c=0.2\cdot n, \ (n=1, 2, \ldots 9)$. The  curves  that have  pronounced half 
squares at the higher part of the figure 
fit the lower values of $c$ and those that have short and flat half squares  
at the lower part belong to the
higher $c$'s.  The horizontal line at the bottom  corresponds to  the 
constant value of the energy obtained at $E \approx V$. Each  quasi-periodic 
half square  is connected at its two sides by
short horizontal linear sections.  One may clearly see that  
the energy is discontinuous and undifferentiable at the vertical sides  
 of each half square.}
\end{figure}

 \begin{figure}
     \includegraphics{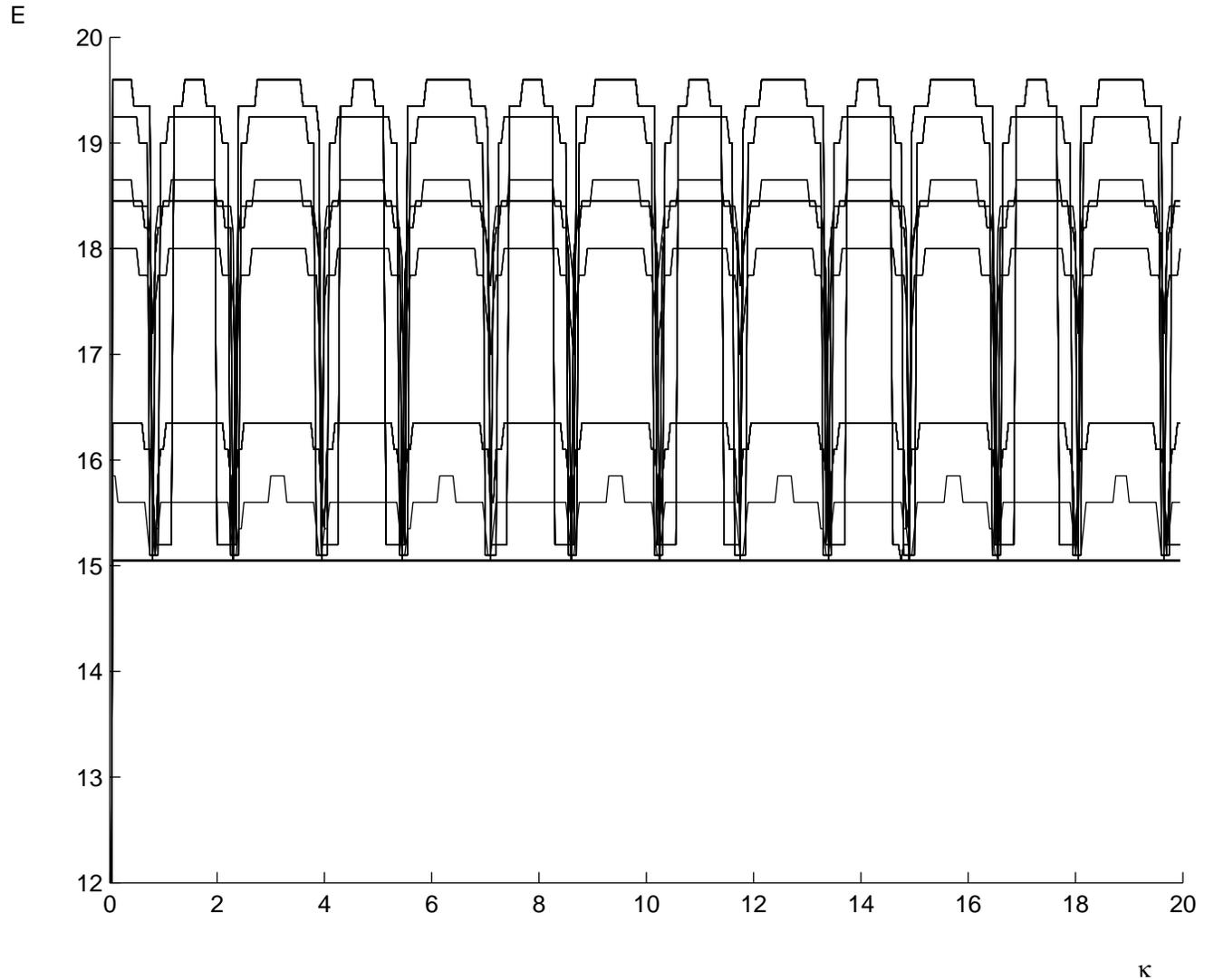}
\caption{The nine graphs in this figure, which show the energy $E$ from Eq
(\ref{e14}) as function of
$\kappa$,   are drawn under exactly the same
conditions as those of Figure 1 except that now $L=5$. One may 
see that the
graphs  are now  separated from each other, especially  those
  at the
higher part of the figure which  are clearly detached from the 
lower three 
 graphs. Note that each half square, especially at the higher part of 
 the figure, is
 discontinuous at several points and not  only at its two sides 
 (compare with 
  Figure 1). 
 The horizontal line at the bottom corresponds to the constant value 
 $E \approx V$. } 
\end{figure}

 \begin{figure}
     \includegraphics{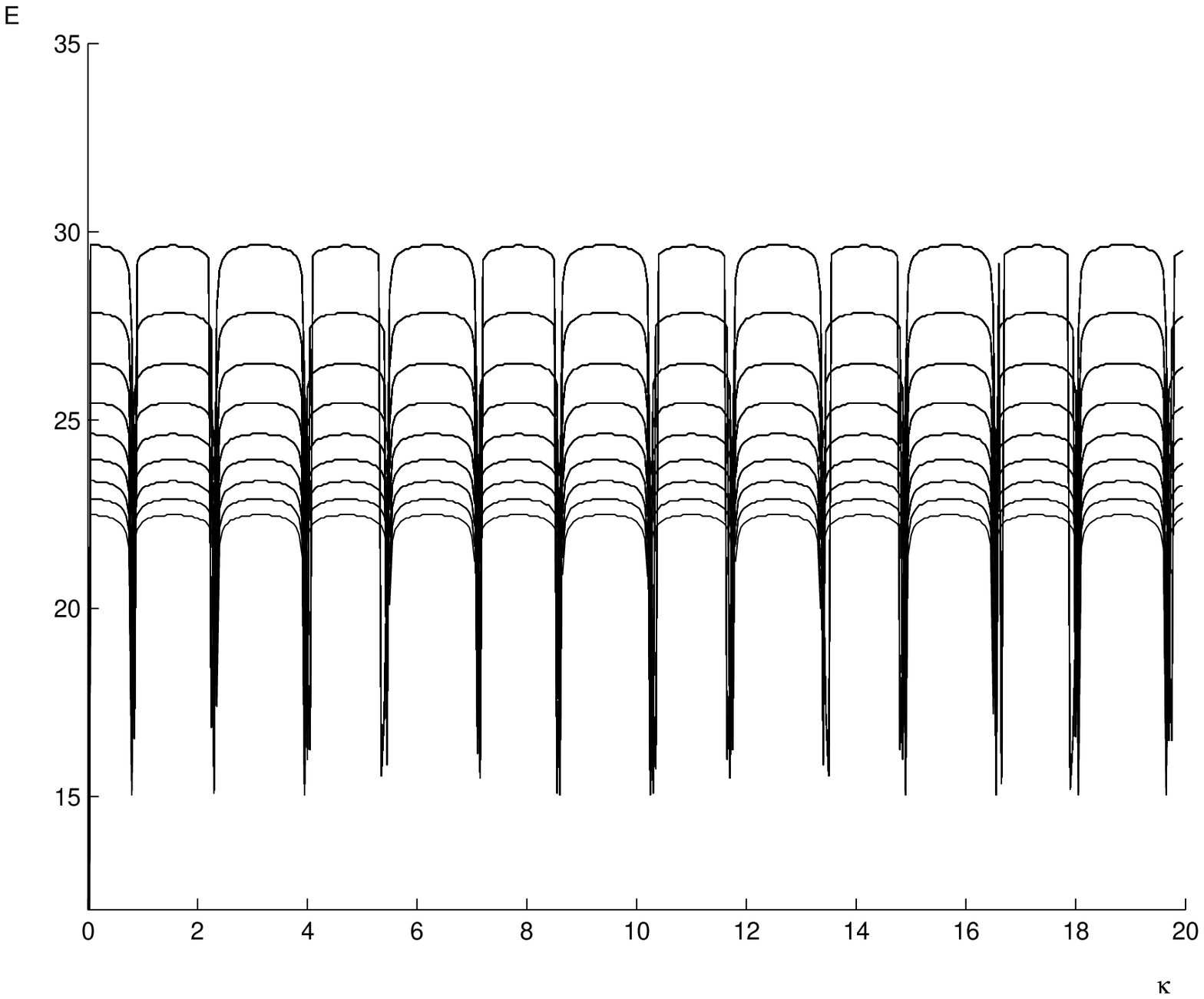}
\caption{In this figure,  which is composed as in Figures 1-2  of 
nine graphs of
the energy $E$ as function of $\kappa$,   the total length $L$ is 
further decreased for
all the  graphs  to  $L=0.3$. The values  of $V$ and
$c$ remain as in Figures 1-2. Note that  this additional  
decrease of $L$ causes
the almost periodic half squares  to be narrower, more curved and arranged 
separately from each  other
compared to Figures 1-2. } 
\end{figure}

  \protect \section{The band-gap structure of the finite multibarrier system 
   for the $E\!<\!V$ case\label{sec4}} 
   We, now,  discuss the $E\!<\!V$ case and begin from Eq (\ref{e12}) which is
   valid, as remarked, also for this case.  We first find    the 
    allowed energies which correspond to the values  of $\kappa$ that cause the left
   hand side of  Eq (\ref{e12}) to vanish, that is, to  
   $\kappa=\pm \frac{(2N+1)\pi}{2}$, $(N=0, 1, 2, 3, \ldots)$.  As for 
   the $E\!>\!V$ case 
these energies are 
   $E=\pm \frac{(2N+1)\pi}{2L^2}+\frac{V}{1+c}$ but  now  
 we  consider  also
   the  $E$'s which correspond to negative values of $\kappa$ so long as
   this does not cause the energies to become negative. That is, considering the
   negative values of $\kappa$ we obtain from the last inline equation for $E$ 
   and  from  
   $E>0$ that $L$, $c$ and $V$ should be related by the inequality 
   $\frac{(2N+1)\pi}{2L^2}<\frac{V}{1+c}$ in which case $L$ could not be very
   small. Thus, taking into account both positive and negative values of
   $\kappa$ and  the condition      
     $E\!<\!V$   we  see that there exist  allowed energies   only 
     for those 
    values of $L$ and $c$  which satisfy  \begin{equation} 
   \label{e19} \pm \frac{(2N+1)(1+c)\pi}{2cL^2} <V, \ \ \ \ \ (N=0, 1, 2, 
   \ldots)  
    \end{equation}  In order to  satisfy  the last inequality for the case in
    which the minus sign at the left hand side is considered the
    parameters $L$, $c$ and $V$ must also be related,  as remarked, by  
    $\frac{(2N+1)\pi}{2L^2}<\frac{V}{1+c}$. Comparing  Inequality (\ref{e19})
     with that of (\ref{e13}) we see,     as for the $E\!>\!V$ case, 
   that  the allowed energies which  correspond to 
    $\kappa=\pm \frac{(2N+1)\pi}{2}$, $(N=0, 1, 2, 3, \ldots)$ 
     critically depend upon the values of the total length $L$ and the 
    ratio $c$. 
    Note that  whereas in  the 
     $E\!>\!V$ case  there are no allowed energies 
     for the very large values of either $L$ or $c$ (or both, see Inequality 
     (\ref{e13})), here, when considering the plus sign at the left hand side of
     (\ref{e19}),  we find no such allowed 
     energies for the  very small 
    values of either $L$ or $c$ (or both).  \par 
    We, now,   find  the allowed energies  that 
    correspond to the values  of $\kappa$ that cause the left hand side of Eq 
    (\ref{e14}) to vanish, that is, to  $\kappa=\pm N\pi$, 
    ($N=0, 1, 2, 3 \ldots$). 
      These  energies are 
found, as for the $E\!>\!V$ case, 
     from $L^2(E-\frac{V}{1+c})=\pm N\pi$ and 
    are  $E=\pm \frac{N\pi}{L^2}+\frac{V}{1+c}$. Using  the former
    discussion at the beginning of this section 
we  conclude that  we may  consider also  negative 
 $\kappa$'s  so
    long as $\frac{N\pi}{L^2}<\frac{V}{1+c}$ in which  case $L$ could not 
    be very
    small. Thus, considering  both positive and negative values of
    $\kappa$ and the  $E\!<\!V$ condition    
      we find  that there are allowed
     energies that correspond to $\kappa=\pm N\pi$, ($N=0, 1, 2, 3 \ldots$) 
     only for  $L$ and  $c$
     which  satisfy  
  \begin{equation} \label{e20}
  \pm \frac{N(1+c)\pi }{cL^2} < V,  \ \ \ (N=0, 1, 2, \ldots) \end{equation} 
  In order to  satisfy  the last inequality for the case in
    which the minus sign at its  left hand side is considered 
  the parameters
  $L$, $c$ and $V$ must be related also by $\frac{N\pi}{L^2}<\frac{V}{1+c}$.  
  Comparing Inequality (\ref{e20}) with that of (\ref{e15}), which refers to the
  $E\!>\!V$ case, we find a result analogous to that   formerly found 
  from  comparing the Inequalities
  (\ref{e13}) and (\ref{e19}). That is,   considering the
  plus sign at the left hand side of (\ref{e20}), one may  realize 
   that   for very 
  small values of either $L$ or $c$ or both of them   there are no allowed 
  energies that correspond  to $\kappa=\pm N\pi, \ \ \ (N=0, 1, 2, \ldots)$. 
      This is  to be compared  to  the corresponding $E\!>\!V$ case and the 
      Inequality
  (\ref{e15}) from  which 
  one  finds that there are no allowed energies for the very large values 
  of either $L$ or $c$ or both of them. \par    
    As
  realized  from Eq (\ref{e14}),  which is valid also for the $E\!<\!V$ case, the dependence 
of the allowed energies   upon  
  $L$ and  $c$  is not restricted  only to these values that correspond to 
  $\kappa=\pm \frac{(2N+1)\pi}{2}$ or to 
   $\kappa=\pm N\pi$, ($N=0, 1, 2, 3 \ldots$). 
Thus, one may find also  the energies
   that correspond to other values of $\kappa$  but in this case 
   we have  first to exclude  those values of 
$E$ that cause the expression that multiply $\tan^2(L^2(E-\frac{V}{1+c}))$ on the right 
hand side of Eq (\ref{e14}) to become negative.  Otherwise, the right hand side of 
Eq (\ref{e14}) would be negative whereas its left hand side is positive. Thus, 
the allowed energies are those that satisfy 
\begin{equation} \label{e21} \frac{V^2}{4E(1+c)(E(1+c)-V)}>-1 \end{equation}
The last inequality results in  
\begin{equation}  \label{e22} E^2-E\frac{V}{1+c}+\frac{V^2}{4(1+c)^2} > 0, 
\end{equation}  
from which one  may conclude that the allowed energies of  the 
bounded multibarrier system
for the $E\!<\!V$ case should be only those which satisfy the inequality 
\begin{equation}  \label{e23} V>E>\frac{V}{1+c} \end{equation}
Thus,  the energies
   that correspond to  $\kappa \ne \pm \frac{(2N+1)\pi}{2}$ and $\kappa \ne \pm
   N\pi$ should belong to the range 
   $V>E>\frac{V}{1+c}$ which means that for very small values of $c$ there are
   no allowed values of $E$ at all. For not very small $c$ 
   the energies  are found by  discussing, as in the $E\!>\!V$ case,
   the two cases of large and small  $c$. That is,   for values of $E$ 
   that are close to $V$ (but always
   $E\!<\!V$) the expression which multiplies $\tan^2(L^2(E-\frac{V}{1+c}))$  
   on the right hand side of Eq (\ref{e14}) 
 tends to
   unity for $c>3$. In such case we may use, as in the $E\!>\!V$ case, the simpler
   Eq (\ref{e16}) which results with  the expression (\ref{e17}) for the allowed 
   energies. 
But now since, as remarked, $E$  
satisfies $V>E>\frac{V}{1+c}$  there are corresponding energies only for 
 $0<N\pi \pm \kappa<\frac{vL^2c}{1+c}$, ($N=0, 1, 2, \ldots$). 
Thus, using Eq (\ref{e17}), 
which is valid also for the
   $E\!<\!V$ case, we find that the  energies which correspond to $c>3$ and
   which are close  from below to $V$ are allowed only for $\kappa$, $c$,
   $L$ and $V$ that are related by \begin{equation} \label{e24} 
   \frac{(1+c)(N\pi \pm \kappa)}{L^2c}<V \end{equation}
   For  values of $c$ from the range $0<c<3$ and for energies that are not
   close  to  $V$ we use the full equation (\ref{e14})
   which necessitates numerical methods for solving  it for $E$ 
    as function of $\kappa$. It is found, as in the $E\!>\!V$ case,  that 
     the required energies are either quasi-periodic or constant. The constant 
     energies
    are demonstrated   by  
    horizontal lines and appear when $E$ assumes the value of $E \approx
    \frac{V}{1+c}$. In this case the right hand side of Eq (\ref{e14}) assumes
    the indeterminate form  $\frac{0}{0}$ so   using L'hospital theorem
    \cite{Pipes} it becomes unity  for all values of $\kappa$. The analogous cases
    discussed in the former section for  $E\!>\!V$  have resulted in finding
    constant horizontal lines for $E \approx v$. \par  Each of the following three
    Figures 4-6 shows 14 different graphs of the energy $E$ as function of
    $\kappa$  for the same value of $v=15$ and 
    for $c=0.2\cdot
    n, \ \ n=1, 2, 3, \ldots 14$.   As for  Figures
    1-3 the 14 curves in  each figure of the set  4-6  fit the 14 values of $c$ 
    in an 
    inverted  order. That is, the highest value of $c$ fits the lowest curve in
    each figure and the second value of $c$ from above corresponds to the second
    curve from below in each figure and so forth. The 14 graphs of Figure 4  are
    all drawn for the  value of $L=30$ and one may see how the energy, in the
    form of half squares, changes almost periodically with $\kappa$. As for the
    $E>V$ case the forms of the energy are not exactly periodic along the
    $\kappa$ axis.   
    One may also  realize that the most
     pronounced half squares  are obtained for the smallest value
    assumed here for $c$ ($c=0.2$). Each half square  is connected 
    at its two sides to its neighbours by
    horizontal lines where,  as for the $E\!>\!V$ case, 
      the energy jumps at the connecting points 
     in a discontinuous manner. Note that the larger is $c$ at  the lower part of
     the figure 
     the longer become the connecting horizontal lines between the half squares 
      which
     become shorter and narrower. \par The  14  different  graphs of the
     energy $E$  as function of $\kappa$, which are shown in Figure 5,  
      are all drawn for the same values 
     of $V$ and $c$ as those of Figure 4 but for $L=5$. The  squared 
     sections   are
     now less emphasized  compared to those of Figure 4 and they become flattened
     for the larger values of $c$ as seen in  the curves at  the lower part of
     the figure. Also, compared to 
      Figure 4,
     each squared section   is connected, especially at the higher part 
     of  Figure 5, 
      to more than two points at which the energy jumps
     discontinuously.  From Figures 4-5 and from the occurence of $L^2$ under
     the {\it tangent} function in Eq (\ref{e14}) one may realize that 
     decreasing the value of $L$ results in  a corresponding
     decrease of the height and width of the quasi-periodic  squared sections. 
     This is clearly
     shown in Figure 6 which shows 14 graphs of $E$ as function of $\kappa$  
     that are  drawn under exactly
     the same conditions and the same values of $V$ and $c$  as 
     those of Figures 4-5 except that now $L$ assumes
     the rather small value of $L=0.8$. The quasi-periodic parts, which 
      are no longer
     half squares,   have been considerably flattened
     to the degree of becoming almost horizontal lines 
      as seen  clearly   in the curves at the lower part of the figure.\par 
      Continuing to further decrease the parameter $L$ results in widening the
      gaps in the energy spectrum, that is, in finding no corresponding allowed
      energies for entire ranges of  the $\kappa$ axis. This is clearly shown in
      Figure 7 which shows the form of $E$ as  function of $\kappa$ for the
      same $V=15$ and the same values of $c=0.2\cdot  n, \ \ n=1, 2, 3, \ldots 14$  as those of
      Figures 1-6 but  for $L=0.278$.   As seen from the figure  
	the allowed energies are   represented by   
	 vertical
	lines between $E=5$ and $E=15$.  These lines differ by their lengths
	which correspond to the different values of $c$. The thick lines signify
    that several lines which correspond to different values of $c$ are
	identical and are seen tangent to each other.   Note also that the
	vertical lines tend to be collected together (see, for example, the
	collection of the 14 lines at $\kappa \approx 20$) and this gathering is
	repeated almost periodically along the $\kappa$ axis where there exists 
	no energy between any two neighbouring quasi-periods. That is, 
	  the  gaps between
	the allowed energies   are quasi-periodically repeated over the  
	positive  $\kappa$ axis 
	 and
	do not vanish  for any value of it.  This result denotes, as will be
	shown in the next section, that the bounded multibarrier potential is
	singular for this value of $L$.  
   \begin{figure}
 \includegraphics{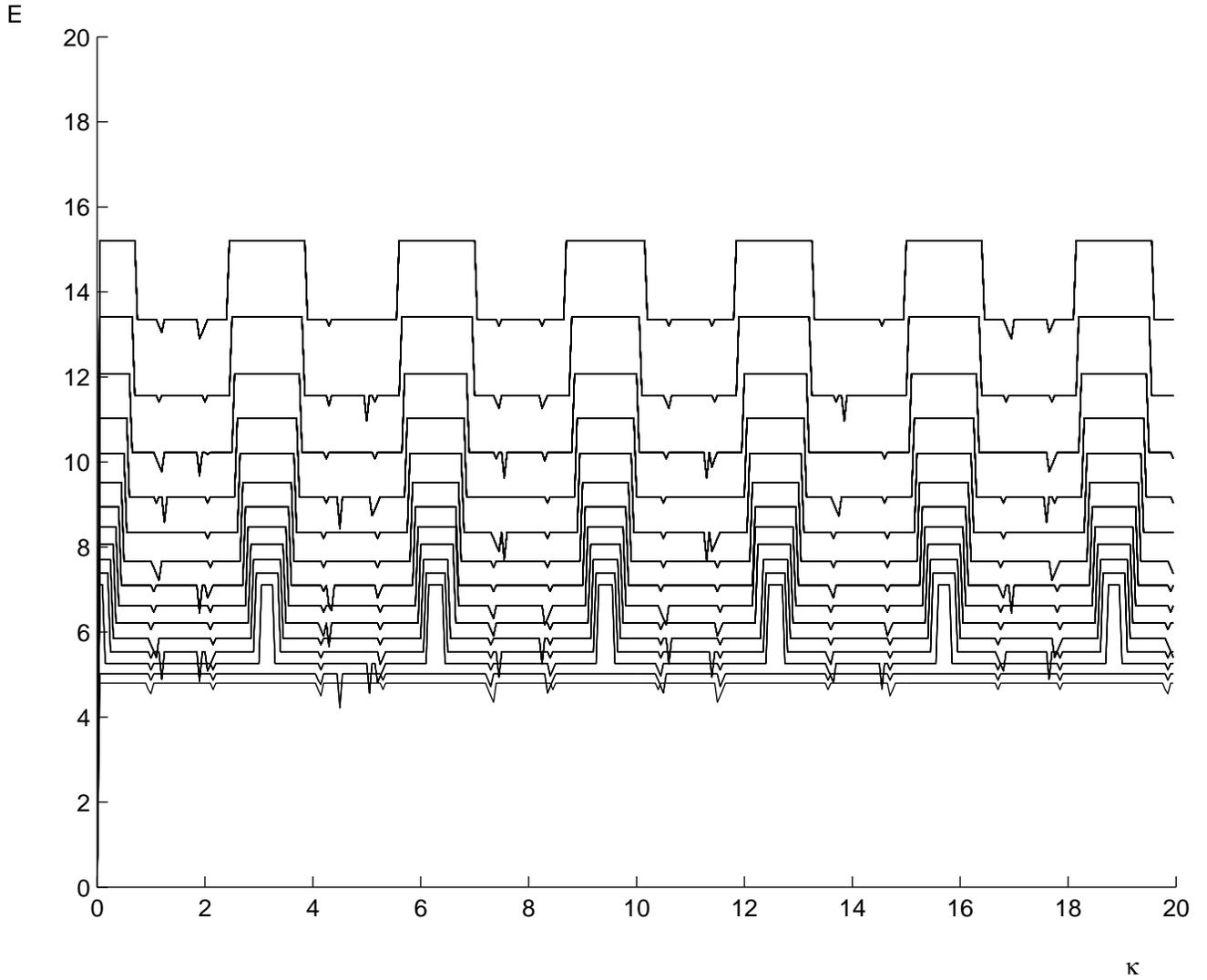}
\caption{The 14 graphs of this figure show the   energy $E$ from Eq (\ref{e14}) 
 as function of
$\kappa$ for the same  $V=15$ and $L=30$ but 14 different values of
$c=0.2\cdot n, (n=1, 2, \ldots 14)$. The allowed energies are found 
in the range
$\frac{V}{1+c}<E\!<\!V$. The graphs fit the values of $c$ in an inverted  
order, 
that is,  the  higher  $c$'s correspond to  the lower curves of the 
figure and vice versa.  
Note that for this value of $L=30$ the almost periodic half squares  are
large and separated   for the smaller  $c$'s  at the higher
part of the figure while  they
become narrow, flat  and mixed among themselves for the larger $c$'s at the lower part of it.} 
\end{figure}

\begin{figure}
 \includegraphics{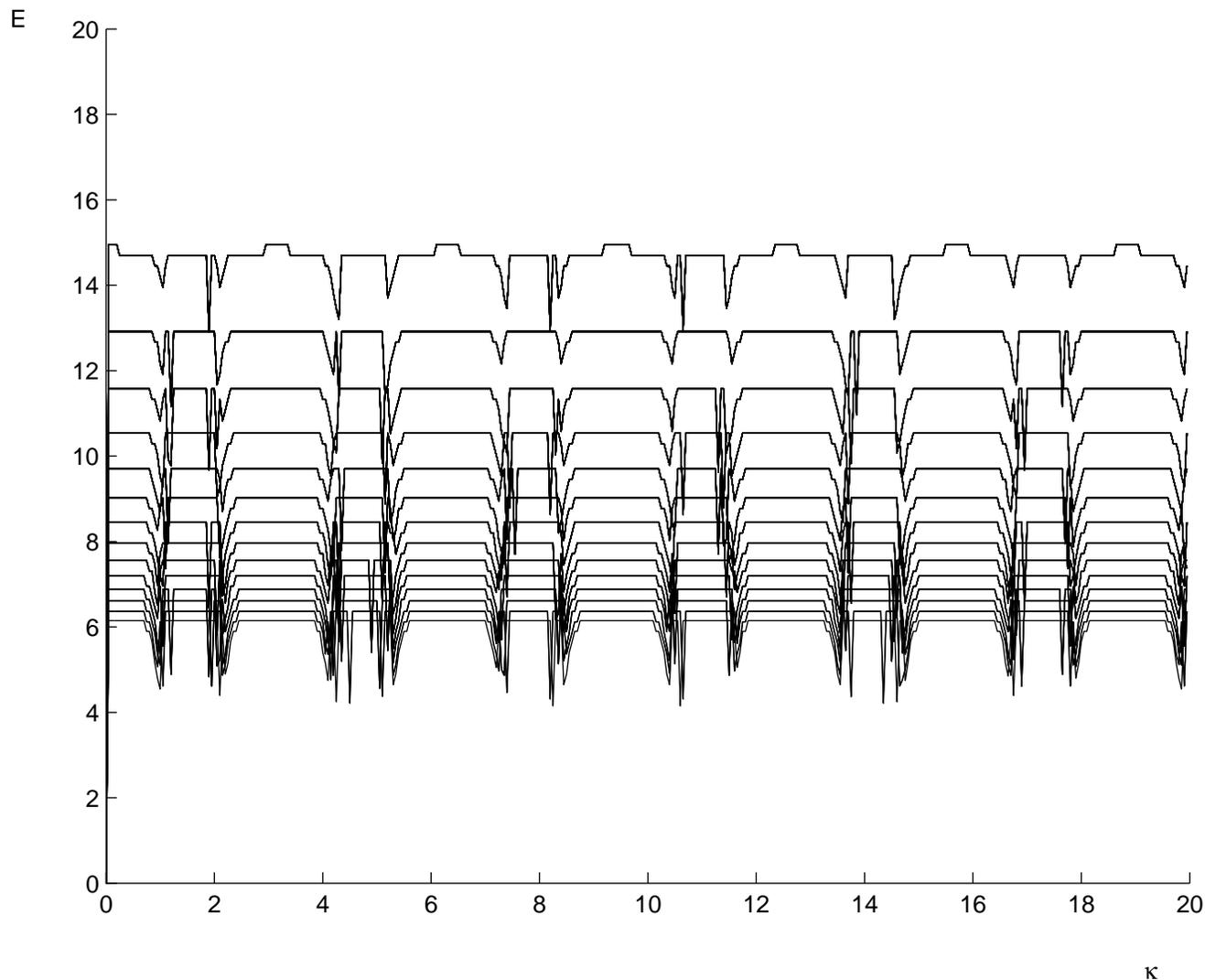}
\caption[fig5]{ All the 14 graphs of this figure, which  show the energy 
$E$ from Eq (\ref{e14}) as function
of $\kappa$,   are drawn for  the same 
$V=15$ and the same values  of $c=0.2\cdot n, (n=1, 2, \ldots 14)$ as those of 
Figure 4 but for the lower value of $L=5$. 
As for Figure 4 the curves  fit
the values of $c$ in descending order so that  the lower values of $c$
correspond to the curves at 
the higher part of the figure and the higher $c$'s fit the curves at 
 the lower part
of it. Note that  the  quasi-periodic  squared sections 
 are less pronounced compared to  those of Figure
4 and  they become almost horizontal lines  for the higher  $c$'s   at 
the lower graphs.   }
\end{figure}

 \begin{figure}
 \includegraphics{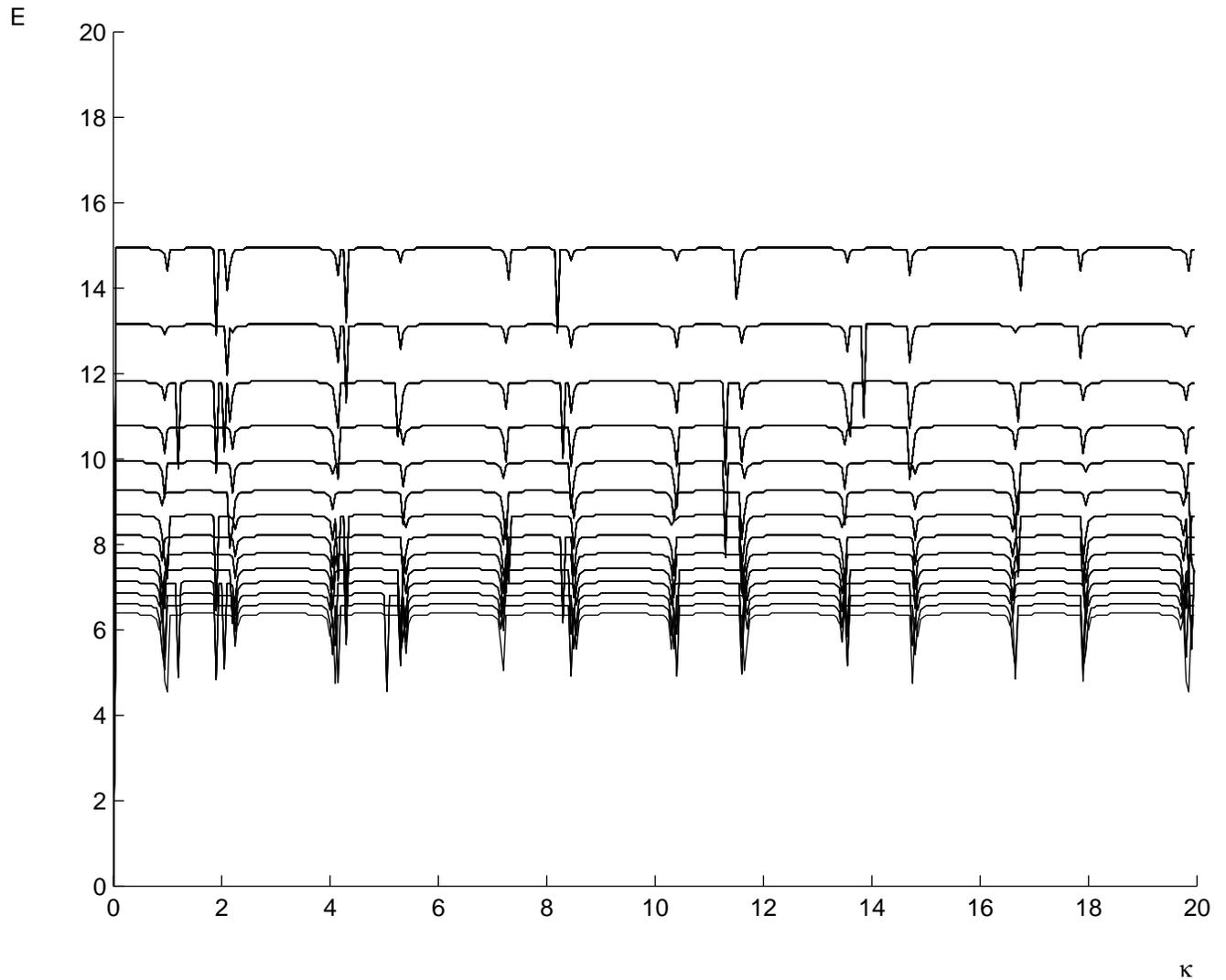}
 \caption{The 14 graphs of this  figure, which show the energy $E$ as function
 of $\kappa$,  are  drawn under exactly the same
conditions and  the same  $V$ and  $c$'s  as those 
 of Figures 4-5 except
that now $L$ is  decreased to the value of  $L=0.8$. 
Note that   the curves become  almost horizontal lines 
and  this  is further emphasized  for the higher values of $c$ 
 in  the lower graphs of the figure.   Also,  one may see that the
 discontinuity encountered for the larger values of $L$ in Figures 4-5 
 is now less pronounced. } 
\end{figure}

\begin{figure}
 \includegraphics{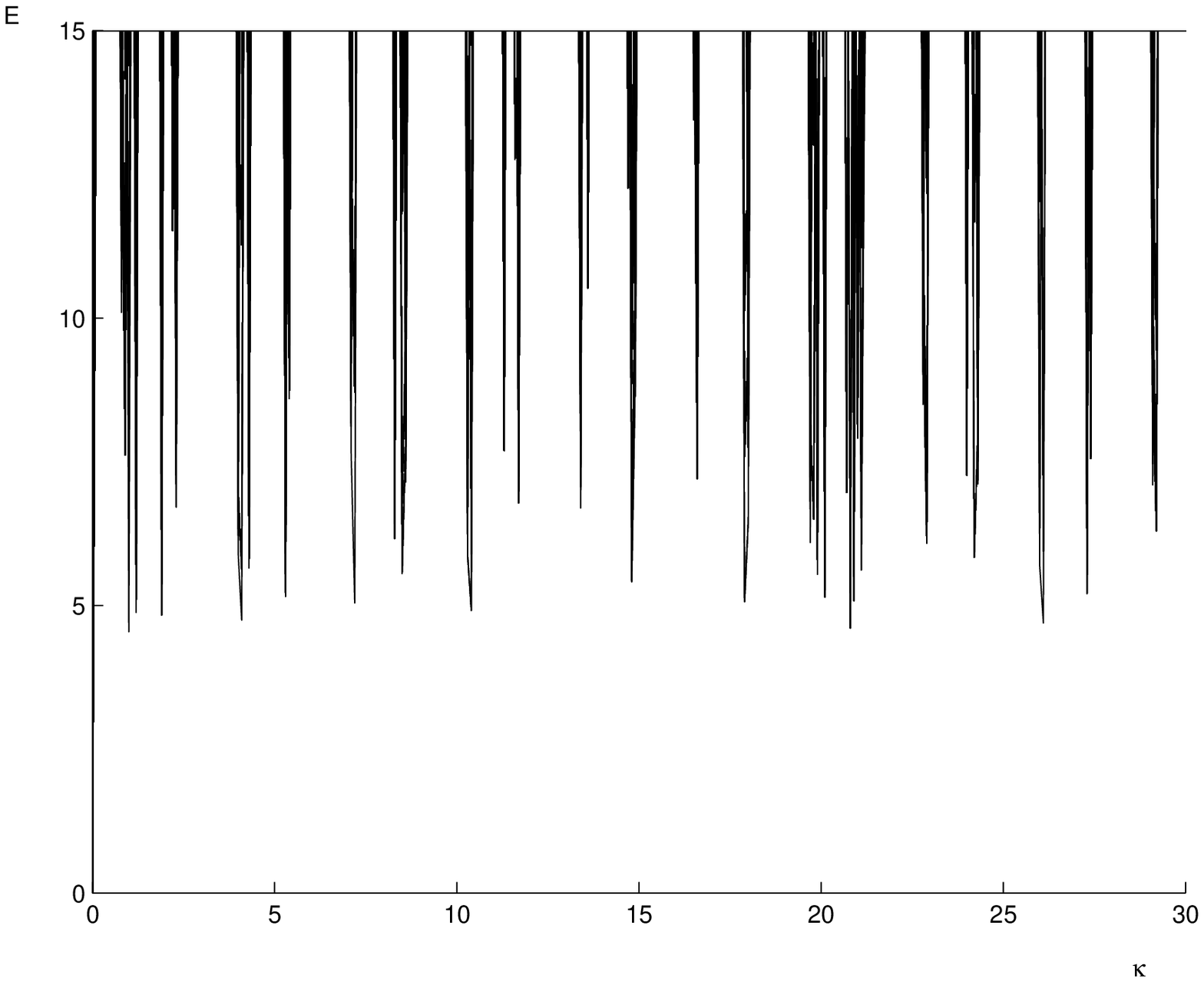}
 \caption{This figure  shows the form of the energy $E$, for the  $E<V$
 case  as function
 of $\kappa$,    for  the same  $V=15$ and the same values of 
 $c=0.2\cdot  n, \ \ n=1, 2, 3, \ldots 14$  as those 
 of Figures 4-6 except
that now $L$ is further decreased to the value of  $L=0.278$. 
As seen,  the allowed energies are represented by the
  vertical lines between $E=5$ and $E=15$  where the thick lines signify that 
  several lines are identical and are drawn tangent to each other. 
  One may realize that  the
gaps for this small $L$ are   repeated over the whole positive 
$\kappa$ axis 
and   do not vanish 
 for any value of it.}
\end{figure}

In Table 1 we show  the allowed energy $E$  and its dependence upon
$\kappa$, $L$,  $c$ and $V$ for both cases of $E\!>\!V$ and $E\!<\!V$. 
 \begin{table}[ht]
   \caption{\label{table1}The table shows the allowed energy $E$  for specific
   values of $\kappa$ and for both cases of $E\!>\!V$ and $E\!<\!V$. We may realize
   from the table that the allowed energies of the bounded multibarrier
   potential depend critically upon the total length $L$ and
   the ratio $c$. Thus, one may conclude  that 
   for certain values of $L$ and $c$ there are no
   corresponding energies as discussed in details in Sections 3 and 4. 
   Also,   there are no  energies, especially, in the 
   limits of very small $L$ or $c$ (or both) for the $E\!<\!V$ case and in the
   limits of  very 
   large  $L$  or $c$ (or both) for the $E\!>\!V$ case.  By $N$ we mean all
   the natural numbers and zero ($N=0, 1, 2, 3, \ldots$).}
\begin{center}
      \begin{tabular}{|c|c|c|} \hline 
      $\kappa$, \ \ $c$ &{\Large $E\!>\!V$}  &
      {\Large $E\!<\!V$} \\ 
      \hline \hline 
      For \ \ all \ \ $\kappa$ \ and  \ small \ $c$ & $E$ \  constant  \ for  \
      $E
      \approx V$ &
     $E $ \ constant  \ for  \ $E \approx \frac{V}{1+c}$ \\
     
       $\kappa=\frac{(2N+1)\pi}{2}$  & $E=\frac{(2N+1)\pi}{2L^2}+\frac{V}{1+c}$ &
     $E=\frac{(2N+1)\pi}{2L^2}+\frac{V}{1+c}$ \\

$\kappa=-\frac{(2N+1)\pi}{2}$  & $E=\frac{(2N+1)\pi}{2L^2}+\frac{V}{1+c}$  &
    $E= -\frac{(2N+1)\pi}{2L^2}+\frac{V}{1+c}$ \\

 $\kappa=N\pi$  & $e= \frac{N\pi}{L^2}+\frac{V}{1+c}$  &
      $E= \frac{N\pi}{L^2}+\frac{V}{1+c}$ \\

 $\kappa=-N\pi$  & $e=\frac{N\pi}{L^2}+\frac{V}{1+c}$  &
     $E=-\frac{N\pi}{L^2}+\frac{V}{1+c}$  \\

    $ N\pi \pm \kappa$, \ $\kappa \neq \pm N\pi$,  \   $\kappa \neq \pm
     \frac{(2N+1)\pi}{2}$, \   $c>3$,  & $E=\frac{N\pi \pm
     \kappa}{L^2}+\frac{V}{1+c}$
       &
    $\frac{V}{1+c}<E=\frac{N\pi \pm
     \kappa}{L^2}+\frac{V}{1+c} <V$ \\

     $N\pi \pm \kappa$, \ $\kappa \neq \pm N\pi$,  \   $\kappa \neq \pm
     \frac{(2N+1)\pi}{2}$, \   $0<c<3$,  & 
     (quasi-periodic,\ \ see \ text)
      &
     (quasi-periodic, \  \ see \ text) \\
 \hline
\end{tabular}
\end{center}
\end{table}

   \section{The singular character of the bounded multibarrier system}
  
 We have seen in the former Sections 3-4 that the bounded multibarrier potential
 have gaps associated with its energy spectrum which depend upon the total length
 $L$ of the system and the ratio $c$ of its total interval to total width.  We
 also show that, unlike the band-gap structure of the infinite Kronig-Penney
 system \cite{Merzbacher} in which the gaps  disappear for large
 values of the energy,   the remarked dependence here of the energy spectrum upon
 $L$ and $c$ implied that for certain values of them the gaps do not disappear. 
  This has been shown for the $E\!>\!V$ case in Section 3 and especially
 for $E\!<\!V$ in Section 4 where we see that there are no energies that satisfy 
 $E \le \frac{V}{(1+c)}$ (see Inequality (\ref{e23}) and also Figure 7 from
 which we realize that there exist quasi-periodic gaps for small values of $L$). 
 Now, it is 
 accepted in the literature
 (see, for example, \cite{Pearson,Avron}) that if  
  the energy spectrum of any
 system have gaps which  do not diminish for either large values of the energy  
 $E$ 
 or large $\kappa$  then the
 relevant system is considered singular. We  have shown  in Sections 3-4 that 
 not only the energy spectrum 
 of the bounded multibarrier potential  have gaps which remain constant
 along the   $\kappa$  axis but that even for the parts of the epectrum which
 have bands there are points at which the energy, as function of $\kappa$, is
 discontinuous. These points are shown,  for both cases of $E\!>\!V$  
 and $E\!<\!V$, to
 be associated with  large $L$ and small $c$  as seen in  
  the higher parts of Figures 1-2 and 4-5. 
  Only at  small 
 $L$  one may obtain energies that change continuously with $\kappa$ and do not
 jump in a discontinuous manner (see Figures 3 and 6). Thus, the existence 
 of the points at  which
 the energy changes discontinuously with $\kappa$  together with the remarked
 nonvanishing gaps demonstrate  that the bounded 
 multibarrier potential is singular.
 This may be corroborated by a  previously obtained result  \cite{Bar1,Bar2} 
    which shows  that the   transmission probability  of the
  bounded one-dimensional multibarrier is unity. 
 That is, the 
 incoming wave function remains the same (preserved) after passing the  dense
 array of potential barriers \cite{Bar1,Bar2}.   Thus, the   three properties which 
 are characterized in the literature
 \cite{Grossmann,Pearson,Avron,Demkov} 
 as signs of singular systems are all found also in  the bounded multibarrier 
 potential discussed
 here. These properties are :  (1) nonvanishing gaps,   (2) the ability to 
  pass all
 the barriers of the system and   
  (3) zero range potential for each barrier which arises here from the fact that
  the bounded spatial length contains very large number of barriers. 
  Thus, we conclude that the dense array of potential barriers is also singular.
   \par  A 
 singular behaviour of the energy spectrum  is 
  demonstrated also in what is
 termed the  singular point interaction $\grave \delta$ which  is  
 characterized by a periodic array of identical short-range perturbations 
 \cite{Grossmann,Avron,Demkov} like 
 our system.  The allowed bands resulting from this $\grave \delta$ 
 interaction are characterized \cite{Avron} by an increasing, as $E$ grows,
  of the gap to band ratio.   
  The  difference between the periodic singular interaction $\grave \delta$
and our system is that the length of the former  is infinite \cite{Avron}, 
whereas our system is
bounded. \par We note that  infinite spatial length   is also what 
characterizes many  examples of singular systems discussed in \cite{Pearson}, 
so the physical
interpretation of the {\it singular continuity}  of these  systems (see 
\cite{Pearson,Avron}) is  connected to their infinity. That is, as written in
\cite{Pearson}   
  {\it "the particle will be transmitted at least
once through each barrier and then it will return again to the 
initial point from which it has started"}. Thus, for these infinite systems 
the singular continuity is characterized by complete reflection that follows a
former stage during  which all the barriers are transmitted. This is so 
for the  infinite systems in which no kind of transmission  
is possible at all, but
since  the bounded system  admits transmission then the particle after passing
all barriers finds itself outside the system at the other side of it and not at
its initial side.  Thus,    what characterizes its singularity  is 
{\it complete transmission after the particle has been transmitted through each 
barrier}.   
This  was clearly shown  \cite{Bar1,Bar2}, using transfer matrix methods
\cite{Merzbacher,Tannoudji,Yu},  for  the  
bounded system discussed here. \par 
 We, now,  show this again by using  the same analytical methods applied  in
\cite{Avron} for showing complete reflection of the $\grave \delta$ system. 
That
is, taking into account the properties of the bounded multibarrier system and
incorporating them into the formalism of \cite{Avron} we show that one obtains
complete transmission instead of complete reflection.  The single change we
introduce into the formalism in \cite{Avron} is that of having to take into
account that there exists only one kind of interaction in the bounded
multibarrier potential compared to   
  the multi-channels \cite{Demkov} version of the  
$\grave \delta$ interaction  discussed in \cite{Avron}. That is, this 
$\grave \delta$  interaction is characterized as being composed from  
all  possible kinds of 
interactions  (elastic, inelastic etc). Thus, 
 the physical system used in
\cite{Avron} is  an infinite sequence of onionlike $N$ 
channels scatterers, so that the reflection amplitude from any such scatterer 
is given in \cite{Avron} by 
\begin{equation} \label{e25} R=\frac{-N^2+1}{N^2+2iN\cot(kl)+1},  \end{equation} 
where $l$ is the length of each scatterer \cite{Avron} and
$k=\sqrt{\frac{2me}{\hbar^2}}$. The total reflection character 
of this system is demonstrated  \cite{Avron}  when 
 the length $l$ tends to zero and the number $N$ of channels  (different
 interactions) in each scatterer 
 becomes very much large.     
  In these limits  one may write \cite{Avron}  $\cot(kl) \approx
  \frac{1}{kl}$ and the unity term in Eq (\ref{e25}) may be  discarded compared
  to $N^2$. Thus,  denoting,  as in \cite{Avron}, the product $Nl$  by 
 $\beta$  
 the reflection amplitude from equation (\ref{e25}) becomes \cite{Avron} 
 $ R=-(1+\frac{2i}{\beta k})^{-1}$. 
 When,  as in \cite{Avron}, $k\to 
 \infty$   (and   still $kl \ll 1$),  the product $\beta
 k$ satisfies, due to the large $N$,  $\beta k=kNl \to \infty$ and the reflection amplitude tends to -1 so
 that the reflection probability $R^2$ becomes $R^2 \to 1$. Now, taking into
 account that for the  single-channel multibarrier potential discussed here
 $N$ is unity we trivially obtain, when substituting  $N=1$ in
  Eq (\ref{e25}), $R=0$. Thus, from the relation between the transmission $T$
  and the reflection $R$ \cite{Merzbacher} $T=1-R$ we learn that $T=1$ as
  obtained from the transfer matrix method in \cite{Bar1,Bar2}. We show in the 
  Appendix that if the total length of the system is bounded then even if the
  interaction is of the multi-channel type,  as  in the $\grave \delta$
  interaction, one obtains zero value for the reflection $R$ provided the number
  of barriers $n$ is much larger than the number $N$ of channels. That is, 
  if the total length $L$ satisfies $L < \infty$ and if $n
  >\!>N$ then $R \to 0$ which means that  $T \to 1$.  \par 
 In summary, we conclude that if the total length of the system is finite 
 then    the singular  character of 
    the  bounded dense
    array of either single or multi-channel scatterers  
    is affected through the complete transmission 
    it demonstrates.

\section{Concluding Remarks}
We have found the allowed energies of the bounded one-dimensional multibarrier
potential for both cases of $E\!>\!V$ and $E\!<\!V$. These allowed energies critically
depend upon the values of the total length $L$ of the system and the ratio $c$
of its total interval to total width.  Thus, it has been shown that there are
values of $L$ and $c$ for which there are no allowed energies. For example, it
has been demonstrated in Section 4 for the $E\!<\!V$ case that the allowed energies
are only those from the range $V>E>\frac{V}{1+c}$. That is, for very small values
of $c$ there exist no energy for the bounded multibarrier system (see also the
constant gaps in Figure 7). \par 
  It has, also, 
been shown for both cases of $E\!>\!V$ and $E\!<\!V$ that for small $c$ and 
large   (or 
intermediate) values of $L$   the energy $E$  have the form of   
quasi-periodic half  squares   
which  are connected to each other by
 horizontal lines.   The points of connection of these half squares, as may be
 seen from the figures,  are points of discontinuity at
 which the energy can not be differentiated.    For small values of $L$ these 
 squared parts  become shorter, narrower and more curved 
 (see Figures 3 and 6) 
 until they disappear for small enough $L$ in which case the energy, as function of
 $\kappa$,  resembles the  form of either  horizontal (or vertical lines 
 for the
 $E<V$ case, see Figure 7). \par
 The remarked existence of gaps which   
  do not disappear  for certain values of $L$ and $c$,      
 even for large  $\kappa$'s,  implied 
  that the spectrum of the bounded multibarrier potential is singular.  
    For example, 
  it has been found for
 the $E<V$ case and for very small $L$ that the allowed energies have the form
 of  vertical lines with nonvanishing  gaps among them as seen in Figure 7. 
 These gaps are quasi-periodically repeated along the $\kappa$ axis 
  and do not vanish for any value of it.  The 
 constancy of  gaps in the energy spectrum of other physical 
 systems,  like the
 $\grave \delta$  mentioned in Section 5, is  interpreted in the
 literature \cite{Avron} as resulting   from  the singularity of the 
 involved systems. \par  We note that the spectrum of the infinite  
  Kronig-Penney periodic potential  is absolutely continuous \cite{Carmona}. In
 such infinite periodic system one can not define, in contrast to bounded
 systems, any total length $L$ or any ratio $c$ of its total interval to 
 total width. That is, the gaps in their energy spectrum, which do not depend on any
 such undefined $L$ and $c$,  can not be preserved, 
 as in the bounded potential,
  merely by taking  some limit or other values of these
  $L$ and $c$. Thus, their spectrum is, as remarked,  generally 
  absolutely continuous
  \cite{Carmona}. 
 There are,  of course, the exceptional infinite periodic systems
 \cite{Pearson,Avron,Demkov} mentioned in Section 5  which are characterized 
 by nonvanishing  gaps 
    but  for other reasons not related at all to any $L$ and $c$.  
    The bounded multibarrier potential discussed here allows one
  not only to define its total length $L$ and the related ratio of its total
  interval to total width but also to express the location of each barrier
  \cite{Bar1,Bar2,Bar3} and consequently  the allowed energies in terms of $L$ and
  $c$. Thus, there may be found, as actually shown in Sections 3-4, specific
  values of $L$ and $c$ for which the gaps in the energy spectrum 
  do not disappear even for large
  values of $E$ or  (and) large $\kappa$.    
 Moreover, for large (and intermediate)  $L$ and small $c$ the energy $E$, 
  as function of $\kappa$,  
 of the bounded 
 multibarrier potential have  
 been shown to have the form of 
 quasi-periodic  half squares  which are jumpy and discontinuous
 at their vertical sides (see Figures 1-2 and 4-5).  
 These  discontinuities  in the  energy  become small and insignificant 
 only when $L$ decreases 
   as may be seen in Figure 3 for the $E\!>\!V$ case
 and in Figure 6 for the $E\!<\!V$ one.  When $L$ becomes very small the gaps in
 the 
 energy
 spectrum becomes wider and are repeated along the positive $\kappa$ axis 
   as seen in  Figure 7. \par 
 In summary,  we see that  the
 bounded one-dimensional multibarrier potential  demonstrates through the  
  discontinuities  
 of $E$, the nonvanishing
 gaps in its  energy spectrum, its almost zero-range potential and    
  its complete transmission   that it is  a  singular system. 

\newpage

 \begin{appendix}
 
 \bigskip \bigskip
 
 \protect \section{ \underline{APPENDIX} \\
 The transmission for the bounded 
 one-dimensional multi-channelled
 multibarrier potential   \label{appen}} 
 
 We, now,  discuss the case of the bounded one-dimensional multi-channelled
 multibarrier potential system and show, using Eq (\ref{e25} and the relation
 $R+T=1$, that the
 transmission  tends in this case to unity      
 The length of each scatterer (barrier) is now  $l=\frac{L}{n}$ where $L$, 
 for the bounded system, signifies  the total length of all the  
    scatterers (barriers), and is some finite number.   
 $n$ is 
 the number of scatterers in the bounded system  and 
 $N$ is the number of channels in each scatterer where we assume  $n \gg N$. 
 We assume in the following that $N$ is large but small compared to $n$ 
 which, actually, satisfies    $n \to \infty$. 
 Thus, denoting as before the product of 
 the length of each
 scatterer and $N$ by $\beta$ (so as to have now $lN=\frac{LN}{n}=\beta$) 
  and using   
  the former approximations of  $\cot(kl) \approx
  \frac{1}{kl}$ and  $N^2+1 \approx N^2$,  we may  
  write Eq
 (\ref{e25})   as \begin{eqnarray}
 R&=&\frac{-N^2+1}{N^2+2iN\cot(\frac{kL}{n})+1}=
  \frac{-N^2+1}{N^2+\frac{2iN^2}{k\beta}+1}  \approx \label{e28} \\ & \approx & \frac{-N^2}
  {N^2(1+\frac{2i}{k\beta})}=-(1+\frac{2i}{k\beta})^{-1} \nonumber  
  \end{eqnarray} Now, when 
  $k\to \infty$ (and still  $kl=\frac{kL}{n} \ll 1$ as for the
   analogous case of  Eq (\ref{e25})) 
    we have now,  due to $n \gg N$,  
  $k\beta =\frac{kLN}{n} \ll 1$, and  the 
  reflection amplitude goes to zero in contrast to the result obtained from Eq
  (\ref{e25}). Thus,  
  from the  relation between the reflection and transmission
  amplitudes    $T=1-R$,  we learn that $T$ goes to 1 as required. That is, 
   we see 
    that if the  length of the sequence of scatterers is  finite  
    and if it is dense enough so that the number of scatterers $n$ is very large
    then 
    the transmission coefficient  goes to 1 even for large number of channels 
    $N$ in
    each scatterer.

     \end{appendix} 
     
     \newpage

 \bigskip \bibliographystyle{}

\end{document}